\documentclass[aps,amsmath,onecolumn,amssymb]{revtex4}
\usepackage{graphicx}
\usepackage{hyperref}
\usepackage{color}

\def\de{\delta}

\def\eps{\varepsilon}
\def\teps{\tilde{\varepsilon}}
\def\th{\theta}
\def\ka{\kappa}

\def\si{\sigma}

\def\Si{\Sigma}
\def\om{\omega}
\def\Om{\Omega}
\def\La{\Lambda}

\def\vphi{\varphi}

\def\na{\nabla}
\def\eti{\eta_{\infty}}

\def\ri{{\rm i}}
\def\rd{{\rm d}}

\def\beq{\begin{equation}}
\def\eeq{\end{equation}}
\def\bea{\begin{eqnarray}}
\def\eea{\end{eqnarray}}
\def\nn{\nonumber}

\def\bR{\mathbf{R}}
\def\bC{\mathbf{C}}
\def\bE{\mathbf{E}}
\def\bS{\mathbf{S}}

\def\Ra{\Rightarrow}

\def\ri{{\rm i}}

\def\diag{{\rm diag}}

\def\bit{\begin{itemize}}
\def\eit{\end{itemize}}

\begin{document}
\title{The Physics of Conformal Cyclic Cosmology}
\author{Krzysztof A. Meissner$^1$ and Roger Penrose$^2$}
\affiliation{$^1$Faculty of Physics, University of Warsaw\\
Pasteura 5, 02-093 Warsaw, Poland\\
$^2$Mathematical Institute, Oxford University,\\
Radcliffe Observatory Quarter,\\
Woodstock Rd., Oxford OX2 6GG, UK
}

\vspace{3mm}

\begin{abstract}
\noindent
According to conformal cyclic cosmology (CCC), the currently conventional description of the entire history of the universe (but without an initial inflationary phase) provides but one cosmic aeon of an unending sequence of such aeons, where the future conformal infinity of each aeon joins essentially smoothly to the conformally stretched big bang of the next, across a spacelike 3-surface, referred to as a crossover 3-surface. Whereas in previous accounts of CCC a detailed description of the physics of crossover had been somewhat problematic, a novel idea is introduced here to show how crossover takes place naturally during a temporal period of the universe that is dominated by gravitational waves referred to here as a gravitational wave epoch (GWE). Accordingly, the geometry at the crossover surface is conformally smooth, except at a discrete set of points, referred to as Hawking points, each representing the final Hawking evaporation of the dominant black hole of a galactic cluster in the earlier aeon. It is shown here (using 2-spinor and twistor techniques) that there is a mass-energy conservation law that holds across the crossover surface, showing that the rise of temperature within such Hawking spots should be effectively determined by the total mass of the pre-crossover galactic cluster involved. This rise of temperature on the CMB map within Hawking spots is found to be in quantitative agreement with the masses of the largest galactic clusters observed in our own aeon what suggests that the physics in the previous aeon was, at least in the gravitational sector, similar to ours. A second observational feature, the actual angular diameter of the Hawking spots seen in our CMB, which is about twice what should have been expected, is associated to the presence of GWE just after the crossover and before the start of the usual cosmological epochs. 

\end{abstract}
\maketitle

\section{Introduction}

According to the Conformal Cyclic Cosmology (CCC) proposal \cite{P1} , the universe consists of a succession of expanding phases referred to as cosmic aeons. Each aeon originates with its own big bang and finally takes part in an indefinitely exponentially expanding de Sitter-like state. The future conformal infinity \cite{P0,PR} of each aeon is thus a spacelike 3-surface, and the aeon is taken to continue, as a conformal-structured space-time, to become the conformally stretched big bang of a subsequent aeon. In the CCC scheme, this succession is considered to continue indefinitely in both temporal directions  \cite{P1,P2,P3}. It is also assumed that each of these aeons is qualitatively similar, in overall features, to each other aeon, all of them expanding, so none actually contracts. Some impressive observational evidence for this picture is given in \cite{AMNP} as well as support provided by \cite{AMN} and \cite{GP}.

This gives us a rough geometrical picture of the CCC model, but we must take this geometry as also providing a plausible picture of the physical situation occurring in our universe. In the first place, it is argued that the CCC model can make sense physically because at both the past and future ends of each aeon, the physics is dominated by conformally invariant physical contributions so that the transition from aeon to aeon can make physical sense through a conformally invariant (scale-free) basic limiting physics, although in this paper the needed conformal invariance takes a different form from that expressed in earlier accounts. 

It should be emphasized that the usually assumed early inflationary phase \cite{Guth} of conventional
cosmology (lasting almost 70 e-folds or more) is necessarily absent in CCC. The various relevant observed effects that in conventional cosmology are explained by an early inflationary phase are explained in CCC by other means, effectively from the fact that the ultimate exponential expansion of the previous aeon behaves, in effect, like an inflationary phase but occurring before, rather than after the big bang of the subsequent aeon. Accordingly, in CCC, there is no horizon problem or isotropy problem, and the presence of a near scale invariance of the temperature fluctuations in the CMB can also be understood, as these are explained in CCC as effects arising from the exponential expansion of the previous aeon, rather than by the imposed very early inflationary phase of currently conventional cosmology. Moreover, anything remotely like the 70 e-fold inflationary phase of conventional cosmology would also be inconsistent with various observed effects \cite{AMN,GP}, as well as \cite{AMNP}, which can be understood as direct implications of CCC.

The common boundary of two successive aeons, connected in the way described here, is referred to as their {\it crossover} 3-surface, this representing both the conformal infinity of the earlier aeon and the conformally stretched big-bang moment of the subsequent one, both being spacelike 3-surfaces \cite{P0,PR}, this being dependent on the assumed positivity of the cosmological constant $\La$, which leads to each aeon having a de Sitter-like remote future.  Accordingly, our own aeon's Big Bang (where the capitalized ``Big Bang'' denotes the origin of the particular aeon that we inhabit) would be considered, when appropriately conformally stretched, to be the conformal continuation of the conformal infinity of a previous aeon, where we denote by $\mathcal X$ the crossover 3-surface joining these two aeons.

Arguments put forward in \cite{P3} suggest that the dark-matter content of an aeon, at a time corresponding to our current moment, would be starting to decay into gravitons, so that eventually -- apart from the effects of $\La$ -- gravitons would dominate the final stages of the aeon. However, since gravitons are quantum particles, one might be tempted to take the view that it would have to be a ''quantum-gravity'' picture governing the transition to the big bang of the following aeon, which might, perhaps, be more in line with much current thinking about the underlying physical nature of the Big Bang. In the CCC picture, on the other hand, the space-time geometry at crossover remains entirely classical though with a conformal, rather than the usual metric structure being smooth at crossover.

To appreciate how a universe phase dominated by quantum entities (gravitons) can remain essentially classical, it is perhaps helpful to compare this situation with that of an ordinary gas, being composed of atoms or molecules which would individually require quantum descriptions; yet we can still take the gas itself to have an adequately classical description. Accordingly, we can take the view that a universe stage which is dominated by gravitons may also have an adequate classical description. Nevertheless, we need to appreciate the kind of space-time geometry that locally represents the purely gravitational degrees of freedom. These may be described by the Weyl (conformal) curvature tensor $C_{abcd}$ -- it actually turns out to have a very simple (and more appropriate) 2-spinor description, which is the same as that for a spin-2 massless quantum particle. Thus, as with the gas considered above, we can indeed take a space-time dominated by gravitons as being described as a classical space-time, which is dominated (apart from the effects of $\La$) by Weyl curvature named below Gravitational Wave Epoch.

Thus, we propose that the conformal infinity of the previous aeon to ours was dominated by gravitational waves in the sense described below. The cosmology described by the Weyl tensor domination is very different from the usual description in terms of the Einstein equation, where a non-vanishing radiation or matter energy-momentum tensor governs the departure of a cosmological model from flat space-time.  Accordingly, we divide the history into phases that we refer to as {\it epochs} to be defined below. Therefore, the history of an aeon would be a sequence of epochs with smooth transitions between them. In the present paper we are concerned mostly with the epoch directly preceding and following the crossover -- we argue that there was an additional period before the crossover, Gravitational Wave Epoch, and short (in conformal time) period in the very early epoch after the crossover needed to explain the length (in conformal time) from the crossover to the Last Scattering Surface that we estimate from observations.

According to CCC, the crossover would provide us with a conformal space-time 4-manifold that is conformally smooth across ${\mathcal X}$ at all points of ${\mathcal X}$, except for a discrete set of points on ${\mathcal X}$ referred to as Hawking points. Each Hawking point would represent the ultimate fate of a galactic cluster in that previous aeon, where we envisage that cluster to have been eventually almost entirely swallowed by its dominant supermassive black hole. In turn, that black hole would, according to standard theory, eventually evaporate away into Hawking radiation. For a cluster of, say, $10^{15} M_\odot$, one might expect a lifetime of around $10^{112}$ years, where, in effect, the galactic cluster's entire mass is eventually radiated away in this form. However, since this period starts extremely late, in the conformal diagram the entire radiated energy is essentially concentrated at a single point H, which is the future end-point of the black-hole's world-line $h$. We refer to H as a {\it Hawking point} which is essentially the final fate of the galactic cluster under consideration, although some proportion of its stars and other material may escape being swallowed by the resulting black hole. From the point of view of the metric geometry of the succeeding aeon, this mass-energy is enormously concentrated at H, and we have to consider that H is a singular point of the conformal geometry, on the crossover surface ${\mathcal X}$, but where we take crossover to be smooth at all points of ${\mathcal X}$ other than Hawking points.

One should mention that one of the predictions of CCC is the presence of a candidate for dark matter \cite{P3} (we actually prefer to use the term 'gravitational matter'). If the gravitational particle, presumably of the Planck scale mass, has the lifetime of the general order of ($\sim 10^{10}$ yrs) then its decay may potentially explain the 'Hubble tension' problem i.e. the difference of the Lema{\^i}tre-Hubble parameter from local measurements and inferred from the CMB observations \cite{PMC} calculated to correspond to the present time. The latter calculation would give a different result from the standard calculation if the matter content changed along the way from the last scattering to the present and it could provide an explanation of the 'Hubble tension' problem. The decay can also be the source of gravitational waves at the end of the aeon leading to the GW Epoch and the subsequent crossover.

As is discussed in this paper the mass transmitted through a Hawking point would have been, in effect, almost the entire mass of the preceding galactic cluster, cannot just disappear in the succeeding aeon -- i.e. into {\it our} aeon -- but we must consider that there is a hugely energetic injection of mass-energy into our own aeon that enters it at the point H. We shall see that this is a consequence of a conservation law that results from the assumed smoothness of the conformal geometry at points of $\mathcal X$ surrounding H where, according to the assumptions of the CCC scheme, we require that the conformal geometry is smooth and, indeed, close to being {\it conformally flat}, at points that are adequately far away from the singular point H. 

In Sections II and III, we show how techniques from 2-spinor and twistor theory allow us to establish the appropriate conservation law and we apply this to provide a means of estimating the mass of the galactic cluster in the previous aeon from the degree to which the temperature is raised, for any chosen 'Hawking spot' as observed in the CMB of our own aeon. As it turns out both values, masses of the largest galactic clusters and the raise of temperature within the spot coming from the total evaporation of the black holes (remnants of these galactic clusters) in the next aeon are consistent with each other. It is very remarkable that these two numbers -- the size of the largest galactic clusters at the end of a previous aeon and the rise of the temperature in the observable Hawking spots -- are consistent with each other.

In a previous article \cite{AMNP}, strong evidence was provided for the existence of spots, of radius around 0.04 radians, having increased temperature in the CMB sky that are not anticipated on the basis of standard cosmology. The existence of spots of this general nature is, however, a prediction of conformal cyclic cosmology (CCC), as explained in \cite{AMNP}. Subsequent theoretical considerations appear to show that these spots, as actually observed, have angular dimensions that turn out to be bigger than the diameter that would have been anticipated on the basis of standard procedures of particle physics and quantum field theory \cite{AG},  as applied to the very early stages of a big bang, in the context of CCC. We calculate what is the required length of an additional epoch in the very early aeon that corresponds to the diameter of these spots as quoted in \cite{AMNP}.

\section{Spinor formalism for conformal space-time geometry}

Let us label our current aeon by $\mathcal A$ and the previous one by $\mathcal B$, the crossover 3-surface between these two aeons being a spacelike 3-surface ${\mathcal X}$, the physical space-time $\mathcal M$ under consideration here being the union of the three
\beq
{\mathcal M}={\mathcal A}\cup {\mathcal X}\cup{\mathcal B}.
 \eeq
 
The world-line of the galactic cluster, under consideration here is the vertical line $h$, within $\mathcal B$, having as its future end-point its {\it Hawking point} H, which is a singular point within $\mathcal X$. The history depicted by $h$ becomes entirely dominated by a single supermassive black hole only for the very top part of $h$, just prior to H itself. This black hole evaporates into Hawking radiation only at a much tinier region, effectively concentrated on the uppermost part of this line, just before terminating at H, all this later activity being compressed to a tiny portion of the line $h$, in effect only infinitesimally below H itself. Within this $\mathcal B$ region (i.e., prior to crossover) we may as well consider that the metric near $h$ is simply that of Schwarzschild, the role of the cosmological constant $\La$ (leading us to consider the Kottler metric) or any residual effects of rotation (leading us to consider the Kerr meric \cite{Kerr}) being unlikely to have significance for us, very near H.

In any case, we need not be much concerned with the full details of this metric here, as we shall be interested, in detail, only with the space-time in regions where the conformal curvature is small, well away from the effectively singular line $h$. Specifically, we are concerned with points close a certain 3-cylinder $\mathcal C$, spatially surrounding $h$, together with its 3-dimensional future ``cap'' $\mathcal H$, which is where the causal future region of H reaches the last-scattering 3-surface $\mathcal L$. In space-time terms, $\mathcal H$ is the ``Hawking spot'' that we are concerned with here. $\mathcal C$ can be represented by two vertical lines in the diagram whose future end-points mark the boundary $\mathcal T$ of $\mathcal H$. The 2-sphere $\mathcal T$ is, indeed, the intersection of the future light cone of H with $\mathcal L$. The cylinder $\mathcal C$ has cross-sections which are spheres $S^2$, the top one in the diagram being $\mathcal C$, which is the boundary of $\mathcal C$'s ``cap'' $\mathcal H$ (where $\mathcal H$ is topologically a 3-ball). The question of where we cut off the cylinder $\mathcal C$ at its {\it past} end $\mathcal P$, (also a sphere $S^2$), is not too important, but we need it to lie within $\mathcal B$, and we take it to share in the spherical rotational symmetry about $h$. We take it that $\mathcal C$ surrounds no significant gravitational sources other than that lying along $h$.

We must bear in mind that a 4-space-time is spherically symmetrical about the line $h$, so that $\mathcal T$ is indeed actually topologically a 2-sphere and $\mathcal H$, a 3-ball (3-disc) under this rotation about $h$. We should also bear in mind that the {\it Hawking spot}, as seen from our vantage point in the universe, would not be the whole of $\mathcal H$, but would only be a 2-dimensional spot, rather than the full 3-dimensional region $\mathcal H$. This reduction in dimension comes about because what we actually ``see'' is only the intersection with our past light cone with what would have been going on in the 3-surface $\mathcal L$ (the ``moment'' of last scattering).

We shall be concerned with calculations within a 4-space region $\mathcal R$ which thickens out the 3-dimensional regions $\mathcal C$ and $\mathcal H$ slightly into the 4-space-time, so that $\mathcal R$ provides a 4-dimensional {\it open-set neighbourhood} of the union  ${\mathcal C}\cup{\mathcal H}$. Moreover, we can consider $\mathcal R$ to have trivial topology (i.e., topology of an open 4-ball), since ${\mathcal C}\cup{\mathcal H}$ also has trivial topology (a 3-ball). The conformal curvature within $\mathcal R$ can be assumed to be reasonably small, deviating only slightly from a portion of a Friedmann-Lema{\^i}tre-Robertson-Walker (FLRW) cosmological model, all of these models being conformally flat, having metrics that can be written in the form
\beq
\rd s^2 =F(\tau)^2(\rd\tau^2-\rd\Si^2),
\label{2.00a}
\eeq
where $\tau$ is what we shall call the {\it conformal time} and where $\rd\Si^2$ stands for the standard squared metric form for Euclidean 3-space, or else for a 3-sphere or hyperbolic 3-space and, within $\mathcal R$, we are nowhere close to the regions where large deviations from FLRW geometry are to be expected, namely near the large-curvature (effectively singular) line $h$, up to its singular end-point H. It should be noted that although the {\it metric} structure of the crossover 3-surface $\mathcal X$ becomes infinite as $\mathcal X$ is approached from above, it is an assumption of CCC that the {\it conformal} geometry is smooth along $\mathcal X$ -- except at Hawking points such as H -- so we can take $\mathcal R$'s conformal structure as remaining smooth along $\mathcal R$, and very close to being conformally flat, well away from H, such as at $\mathcal C$'s future end-region $\mathcal T$ and the 3-ball $\mathcal H$ that bounds it. Accordingly, there should be no obstruction to taking the space-time region $\mathcal R$ to be treated, as a whole, as being a reasonably small perturbation away from a conformally flat space-time region. In addition to this, we may as well consider that the metric form $\rd\Si^2$  is actually that of Euclidean 3-space, since not only are we looking at a very small region of the whole sky (less than one tenth of a radian across), but also because of strong observational limits to the existence of spatial curvature of the whole universe.

We start with the Riemann tensor written as
\beq
R^{ab}{}_{cd}=C^{ab}{}_{cd}+E^{ab}{}_{cd}+\frac16 R\de^{[a}_{c}\de^{b]}_{d}
\eeq
where $C$ is the Weyl tensor and
\beq
E^{ab}{}_{cd}=2\tilde{R}^{[a}_{[c}\de^{b]}_{d]},\ \ \ \tilde{R}^a_c=R^a_c-\frac14 R\de^a_c
\eeq
The sign of the Riemann tensor is taken to be consistent with
\beq
(\na_a \na_b-\na_b \na_a ) V^d =R_{abc}{}^d V^c,
\label{2.0c}
\eeq
where $\na_a$ stands for covariant derivative with respect to the metric tensor $g_{ab}$. 
Under a conformal rescaling of the metric tensor
\beq
g_{ab}\mapsto  \tilde{g}_{ab}=\om^2 g_{ab},\ \ \     g^{ab}\mapsto \tilde{g}^{ab}=\om^{-2} g^{ab}
\label{2.a}
 \eeq
(where $g_a^{b}$ is, in effect, the Kronecker delta $\de_a^b$, and is unaffected by conformal rescaling), we find that the Weyl conformal curvature tensor scales as
\beq
C_{abcd}\mapsto \tilde{C}_{abcd}=\om^2 C_{abcd} \Ra C_{abc}{}^d\mapsto \tilde{C}_{abc}{}^d = C_{abc}{}^d
\label{2.b}
\eeq
The condition, for a local region of space-time (of trivial topology), that it be conformal to a region of Minkowski space-time $\mathbb M$ is that the Weyl tensor vanish throughout that region.

In many of the calculations which follow, it is simplest (and most revealing) to use a 2-spinor formalism and, for this, we shall follow the notation and formalism of \cite{PR}. The 2-spinor indices will be denoted by capital italic Latin letters, {\it primed} when referring to the complex-conjugate spin-space, and the formal index translation from tensor to 2-spinor form is:
\beq
a = AA',\  b = BB',\  c = CC',\   d = DD', \ \ldots
\eeq
(all in abstract-index notation \cite{PR}) and the metric tensor translates to the product of skew-symmetric spinor epsilons according to 
\beq
g_{ab} = \eps_{AB} \eps_{A' B' },\ \ \    g^{ab} = \eps^{AB} \eps^{A' B' },
\label{2.c}
\eeq
where
\beq
\eps^{AB} =-\eps^{BA},\ \ \    \eps_{AB} =  -\eps_{BA},  
\eeq
and analogously for primed indices. $\eps_A{}^{B}= -\eps^{B}{}_ A$  and $\eps_{A'}{}^{B'}= -\eps^{B'}{}_ {A'}$  are Kronecker delta symbols. 

Spinor indices being raised and lowered according to
\beq
X^A = \eps^{AB} X_B, \ \ \  X_B = X^A \eps_{AB}
\label{2.d}
\eeq
and analogously for primed indices. Under the conformal rescaling (\ref{2.a}), we take the (abstract-indexed) epsilon spinors to scale, in accordance with (\ref{2.a}) and (\ref{2.c}), as:
\beq
\eps_{AB}\to\teps_{AB}=\om\eps_{AB},\ \ \    \eps^{AB}\to\teps^{AB}=\om^{-1} \eps^{AB }
\label{2.e}
\eeq
and analogously for primed indices.

In spinor terms, the Weyl tensor takes the form
\beq
C_{abcd} = \Psi_{ABCD} \eps_{A' B' } \eps_{C' D' } + \eps_{AB} \eps_{CD} {\bar\Psi}_{A' B' C' D'},
\label{2.f}
\eeq
where the somewhat complicated symmetry and trace-free conditions of satisfied by $C_{abcd}$ are now replaced simply by the total symmetry of $\Psi_{ABCD}$:
\beq
\Psi_{ABCD} = \Psi_{(ABCD)}.
 \eeq 
Moreover, in vacuum space-time (with cosmological constant $\La$),
\beq
R_{ab} =\La g_{ab},
\label{2.ff}
\eeq
we find \cite{PR} that the Bianchi identity can be expressed as
\beq
\na^{AA' } \Psi_{ABCD} = 0.
\label{2.g}
\eeq
This equation is of particular interest, because it is the wave equation for a massless field of spin 2 (see \cite{Dirac,PR}). More technically correct would be to say that (with the conventions of \cite{PR}) equation (\ref{2.g}) refers to a left-handed, or negative-helicity massless wave-function for spin 2, i.e. for helicity $s=-2$. The helicity $s=+2$ part of the wave-function would be described by the complex-conjugate form of (\ref{2.g}), namely
\beq
\na^{AA'} {\check\Psi}_{A' B' C' D' } = 0.
\label{2.gg}
\eeq
For a real space-time, as opposed to a quantum wave-function, we would have 
\beq
{\check\Psi}_{A' B' C' D'} = {\overline\Psi}_{A' B' C' D'},
\eeq
so that (\ref{2.gg}) gives us nothing new. But if we are indeed thinking in terms of a {\it graviton wave-function}, we can use complex quantities like $\Psi$ and $\check\Psi$, subject, respectively, to (\ref{2.g}) and to (\ref{2.gg}), to describe the negative and positive helicity parts of a graviton's wave-function \cite{PR}.

 We may compare this with the 2-spinor form of Maxwell's equations in free space (i.e., when the charge-current vector $J_a$ vanishes)
\beq
\na^{AA' } \vphi_{AB} = 0.
\label{2.h}
\eeq
(with $\vphi_{AB}=\vphi_{BA}$) where the Maxwell field tensor is
\beq
F_{ab} = \vphi_{AB} \eps_{A' B' } + \eps_{AB} {\overline\vphi}_{A' B' }
\label{2.i}
\eeq
which we may compare with (\ref{2.f}). Again, we can consider wave-functions, this time for photons, for which we would take (\ref{2.i}) to give a splitting into negative ($s=-1$) and positive ($s=+1$) helicities (with the positive helicity $\check\vphi_{A' B' }$ that is independent of the negative helicity $\varpi_{AB}$).   

Here, we are not concerned with wave-functions as such, but the 2-spinor formalism is still enormously helpful, particularly in relation to conformal invariance. For spin $n/2$ (with $n\ne ­0$), in conformally flat space-time $\mathcal M$, we can consider the equation
\beq
\na^{AA' } \phi_{ABC\ldots F} = 0.
\label{2.j}
\eeq
where $\phi_{ABC\ldots F}$ has $n$ symmetric indices, and we may take (\ref{2.j}) to be the wave equation for helicity $s=-n/2$. Equation (\ref{2.j}) is conformally invariant \cite{PR}, that is to say, (\ref{2.j}) retains its form under conformal rescaling, provided that $\phi_{ABC\dots F}$ has conformal weight $-1$, where we say that, under the conformal re-scaling (\ref{2.a}), with (\ref{2.e}), that a quantity $Q_{\ldots}^{\ldots}$ has conformal weight $w$ if it scales as
\beq
Q_{\ldots}^{\ldots} \mapsto Q_{\ldots}^{\ldots} =\om^w Q_{\ldots}^{\ldots}
\label{2.k}
 \eeq
It is of importance to note that although, in vacuum, the Weyl spinor $\Psi_{ABCD}$, as defined in (\ref{2.f}), satisfies the same equation (\ref{2.g}), as that shown in (\ref{2.j}), it has a conformal weight, as determined by (\ref{2.f}), and by (\ref{2.b}) and (\ref{2.e}), which is 0, rather than the $-1$ that would be required for the conformal invariance of (\ref{2.g}). To accommodate this apparent discrepancy, it will be very significant for us to define a quantity $\psi_{ABCD}$ which, in the physical metric $g_{ab}$ of $\mathcal M$, is equal to $\Psi_{ABCD}$:
\beq
\psi_{ABCD}=\Psi_{ABCD},
\label{2.l}
\eeq
but where, after conformal rescaling to the metric $\tilde{g}_{ab}$ ($=\om^2 g_{ab}$) of (\ref{2.a}) we demand
\beq
\tilde{\psi}_{ABCD}=\om^{-1} \tilde{\Psi}_{ABCD}.
\label{2.m}
\eeq
Then $\psi_{ABCD}$ has the required conformal weight $-1$ needed for the conformal invariance of (\ref{2.j}), with $\psi_{ABCD}$ in place of $\phi_{ABCD}$.

There is no inconsistency here, because we would not normally expect the ``vacuum equation'' $\tilde{R}_{ab}=\La \tilde{g}_{ab}$ of (\ref{2.ff}) to hold, with $\tilde{R}_{ab}$ defined in terms of the covariant derivative $\tilde{\na}^a$ determined by $\tilde{g}_{ab}$ (see (\ref{2.0c})), so that $\tilde{\na}^{AA' } \tilde{\Psi}_{ABCD}$ would generally {\it not} vanish, whereas the quantity $\tilde{\psi}_{ABCD}$, with its $-1$ weighting, would still satisfy  
\beq
\tilde{\na}^{AA'} \tilde{\psi}_{ABCD} = 0
\label{2.n}
\eeq
by virtue of the vacuum equations $R_{ab}=\La g_{ab}$ holding for the {\it physical} $g_{ab}$ metric.

In fact, the quantity $\psi_{ABCD}$, with its conformal weighting $-1$, will have an important role for us here, of a slightly different character. We shall take it to represent the Weyl curvature of a {\it weak-field perturbation} away from a conformally flat background. This background here will be that of the cosmology described by (\ref{2.00a}),
namely
\beq
\rd s^2 =F(\tau)^2(\rd\tau^2-d\Si^2),
\label{2.p}
\eeq
where $\rd\Si^2$ is the squared metric form for Euclidean 3-space, so that
\beq
\rd s^2 = \rd\tau^2-\rd\si^2 =\om^2 \rd s^2,
\label{2.pp}
\eeq
provides our metric $\tilde{g}_{ab}$ for Minkowski space-time $\mathbb M$. This is consistent with making the choice
\beq
\om=\pm F(\tau)^{-1},
\label{2.q}
\eeq
in (\ref{2.a}) and (\ref{2.c}), where the $+$ sign refers to region $\mathcal A$ and the $-$ to region $\mathcal B$, as will be explained later.

At this point, some clarification is needed as to how we are treating the physical space-time $\mathcal M$, with is metric $g_{ab}$. We shall be concerned, primarily, with the region $\mathcal R$, described above. Although $\mathcal R$ contains a region of very large curvature just to the future of $\mathcal X$, it nevertheless remains far from any region of $\mathcal M$ where the {\it conformal} curvature gets large, namely along $h$ with its future end at the Hawking point H. Thus, perturbations away from the space-time $\mathcal M$, can still be considered to be small if the {\it conformal} structure across $\mathcal X$ remains smooth. In accordance with the principles of CCC, the crossover region $\mathcal X$ is indeed taken to be conformally very smooth, except at the Hawking points. This will enable us to understand how the mass associated with the line $h$ in the aeon $\mathcal B$, can be transferred into the aeon $\mathcal A$ by investigating how $\psi_{ABCD}$ behaves in regions of $\mathcal X$ {\it surrounding} the singular point H.

The quantity $\tilde{\psi}_{ABCD}$ (with conformal weight $-1$), subject to the conformally invariant propagation equation (\ref{2.m}) in flat space-time $\mathbb M$ (with its flat metric $\tilde{g}_{ab}$) describes the 1${}^{st}$-order small departure from the conformal flatness of $\mathbb M$. The covariant derivative operator $\tilde{\na}_a=\tilde{\na}_{AA'}$ is taken to be that of this (conformally) flat background. We do not need to consider $2^{nd}$-order derivative expressions such as (\ref{2.0c}). Accordingly, we can appeal to studies of this equation in flat space-time $\mathbb M$. Indeed, there is a considerable literature concerning solutions of (\ref{2.n}), in $\mathbb M$, which have essentially as much freedom as complex solution of the wave equation (see (e.g. \cite{PR})). We shall regard such a solution for $\tilde{\psi}_{ABCD}$ as providing us with a specific weak-field perturbation of the metric (\ref{2.p}), away from exact conformal flatness, and providing us with its conformally rescaled weak-field Weyl curvature. 

It should be mentioned that there are various other reasons for taking $\tilde{\psi}_{ABCD}$, rather than  $\tilde{\Psi}_{ABCD}$, for the role, one carrying information from one aeon to the next. The most blatant of these is that, according to the work of Friedrich \cite{Friedrich}, we must expect that, in a generic expanding universe model, with positive $\La$, the conformal curvature ($\tilde{\Psi}_{ABCD}$) will {\it vanish} at its spacelike conformal infinity ${\mathcal I}^+$. The derivative of the conformal curvature away from ${\mathcal I}^+$ will, in general, not vanish, and examining this derivative amounts to the same thing as examining the value of $\tilde{\psi}_{ABCD}$ at ${\mathcal I}^+$.

This is consistent with work done, primarily around 1960 when gravitational radiation in asymptotically flat space-times was being considered \cite{AT}. In order to study gravitational radiation and, in particular, to determine how much the mass of a system is reduced because of the energy being carried away by such radiation, it had proved convenient to apply a conformal rescaling to the space-time to provide it with a future conformal boundary ${\mathcal I}^+$, the relevant quantities becoming finite at ${\mathcal I}^+$. In this work, with a conformal factor $\om$ that smoothly approaches 0 in null directions away from the sources, defining ${\mathcal I}^+$ as where  $\om=0$.

The situation being considered here has much in common with this earlier work, particularly in that the gravitational quantities studied at ${\mathcal I}^+$ tended to be component of $\tilde{\psi}_{ABCD}$ appearing there, the main difference being that in that earlier work, ${\mathcal I}^+$ was taken to be null (corresponding to $\La =0$), rather than spacelike for our current considerations (corresponding to $\La>0$). In each case we expect to obtain a measure of the total mass of a system by performing an integral over a sphere on ${\mathcal I}^+$. We shall consider how to do this in the current situation in the next section.

\section{Conserved mass in CCC}

It is important for our discussion here that we introduce the notion of a {\it twistor} which, in a conformally flat space-time, is a {\it helicity raising operator}, or in the current context, a spin-lowering operator (lowering the spin of a {\it negative}-helicity field by $\frac12\hbar$). For our purposes here, a twistor may be regarded as a solution $\varpi^A$, of the {\it twistor equation}
\beq
\na_{A'}^{(A} \varpi^{B)} = 0,
\label{3.a}
\eeq
which is conformally invariant, with $\varpi^A$ having conformal weight 0 \cite{PR}. This equation has 4 independent solutions in a conformally flat space-time $\mathcal N$, taken to have trivial topology. The solution space of (\ref{3.a}) is thus a 4-dimensional complex vector space $\mathbb T$, referred to as the {\it twistor space} of $\mathcal N$, see \cite{PR}. Each element $\varpi^A$ of $\mathbb T$ (satisfying (\ref{3.a})) serves to transform any solution $\phi_{ABC\ldots EF}$ of (\ref{2.j}), for spin $n\ge 1$ (i.e. with at least two indices) into another solution
\beq
\phi_{ABC\ldots EF} \varpi^F
\label{3.b}
\eeq
of the massless free-field equation (\ref{2.j}), but with one index fewer, as is easily checked.

Let us now apply this procedure {\it twice} in succession to, in particular, a symmetric the 4-indexed quantity $\psi_{ABCD}$, subject to the massless free-field equation (see (\ref{2.g}))
\beq
\na^{AA' } \psi_{ABCD} = 0
\label{3.bb}
\eeq
by using two different solutions $\varpi^C$ and $\widehat{\varpi}^D$ of (\ref{3.a}). Then we find that,
\beq
\phi_{AB}= \psi_{ABCD} \varpi^C \widehat{\varpi}^D
\label{3.c}
\eeq
satisfies the free Maxwell equation
\beq
\na^{AA' } \phi_{AB}=0,
\label{3.d}
\eeq
(see (\ref{2.h})), where we have come down from spin 2 to spin 1 in two steps of $\frac12$. However, it is more economical to lower the spin by 1 in a single step, by using what would be a symmetric rank 2 twistor, rather than two independent rank 1 twistors (each coming from a solution of (\ref{3.a})). The equation for a symmetric rank-2 twistor (sometimes referred to as a ``Killing spinor''; see [\cite{PR} eq(6.7.16), \cite{WP}) is 
\beq
\na_{A'}^{(A} \ka^{BC)} = 0
\label{3.e}
\eeq
where $\ka^{AB}=\ka^{BA}$, which is a conformally invariant equation, $\ka^{AB}$ having conformal weight 0. \cite{Killing}.The solution space of such quantities is $10(=4\times 5/2)$-dimensional, being the symmetric product of the 4-dimensional twistor space $\mathbb T$ with itself \cite{PR}, section 6.4]. Accordingly, we can replace (\ref{3.c}) by
\beq
\phi_{AB}= \psi_{ABCD} \ka^{CD}
\label{3.f}
\eeq
and the free Maxwell equation (\ref{3.d}) again follows from (\ref{3.bb}).

Let us now take our conformally flat space-time $\mathcal N$, under consideration here, to be flat Minkowski 4-space $\mathbb M$, where we are concerned with a weak-field gravitational perturbation away from $\mathbb M$,    described by $\psi_{ABCD}$ (i.e. the weak-field version of the  $\Psi_{ABCD}$ of (\ref{2.g})), in a vacuum sub-region $\mathcal V$ of $\mathbb M$, so that (\ref{3.bb}) holds true within $\mathcal V$. We take $\mathcal V$ to surround some gravitational source that we are interested in. The idea here is that this source will arise as a source of the free Maxwell field $\phi_{AB}$ of (\ref{3.f}), for suitable $\ka^{AB}$. The strength of that electromagnetic source can be calculated by performing an appropriate Gauss integral of $\phi_{AB}$, taken over a closed 2-suface $\mathcal S$, lying within this source-free region $\mathcal V$. This allows us to obtain the total charge, for this field, that is surrounded by $\mathcal S$ -- or more appropriately, in space-time terms, ``threading through'' $\mathcal S$. The idea here is that when we apply this to the $\phi_{AB}$ defined by (\ref{3.f}), using the correctly chosen $\ka^{AB}$ satisfying (\ref{3.e}), we obtain the total mass for $\psi_{ABCD}$ surrounded by $\mathcal S$.

Such an integral could take the form
\beq
\int \phi_{AB} \eps_{A' B' } \rd x^a\rd x^b .
\label{3.g}
\eeq
The exact definition of $x^a$ in this expression is not important, (although we can take it to be the position vector that will be defined shortly in relation to (\ref{3.r}) below), since (\ref{3.g}) simply represents the standard procedure of integrating a differential form (standard ref, for exterior calculus—or \cite{PR}), which is independent of any specifically chosen coordinate frame. Essentially, the integrand is a 2-form, i.e. a skew-symmetric tensor, say $p_{ab} (=p_{[ab]}$ ), which is here given by
\beq
p_{ab}=\phi_{AB} \eps_{A' B' }.
\label{3.h}
\eeq
If this 2-form is closed, i.e.
\beq
\na_{[a} p_{bc]}=0,
\label{3.i}
\eeq
then the integral (\ref{3.g}), of $p_{ab}$, over the closed 2-surface $\mathcal S$ remains constant as $\mathcal S$ is continuously moved over any region where (\ref{3.i}) remains true, as would be necessary if the expression (\ref{3.g}) is to represent some source quantity external to, but surrounded by, the region $\mathcal V$. This is indeed the case here, because (\ref{3.i}) is equivalent to
\beq
\na_a p_{bc} e^{abcd}=0,
\label{3.j}
\eeq
where $e_{abcd}$ is the completely skew-symmetrical dualizing tensor ($e^{abcd}=e^{[abcd]}$ where $e_{0123}=1$). In 2-spinor terms \cite{PR} we find
\beq
e^{abcd}=\ri \eps^{AC} \eps^{BD} \eps^{A' D'} \eps^{B' C'}-\ri \eps^{AD} \eps^{BC} \eps^{A' C'} \eps^{B' D' }
\label{3.k}
\eeq
so that the expression in (\ref{3.j}) indeed vanishes when (\ref{3.d}) holds, because the latter actually asserts that
\beq
\na_A^{A'} \phi_{BC}
\label{3.m}
\eeq
is symmetric in $AB$ and therefore totally symmetric in $ABC$, so that when contracted with (\ref{3.k}) both terms must vanish by the skew symmetry of $\eps^{AC}$ or $\eps^{BC}$ in (\ref{3.k}). It follows that the integral (\ref{3.g}) indeed provides us with the value of a “source” for the surrounding spin-1 free field $\phi_{AB}$ and therefore for the free spin-2 field $\psi_{ABCD}$ for each solution $\ka^{AB}$ of (\ref{3.e}).

It should be noted that the integral expression (\ref{3.g}), as applied to a symmetric spinor field $\phi_{AB}$ (subject to (\ref{3.d})) would in general be a {\it complex} quantity, rather than a real one, since $\phi_{AB}$ is complex. For $\phi_{AB}$ related to the Maxwell field $F_{ab}$ according to (\ref{2.i}), this would allow for the possibility of a {\it magnetic} as well as an electric source for the field $F_{ab}$. Normally, with regard to the expression (\ref{3.g}), we would require that the electromagnetic field $F_{ab}$ be real and derivable from a potential $A_a$:
\beq
F_{ab}=\na_a  A_b -\na_b  A_a ,
\label{3.n}
\eeq
in which case the magnetic charge would be zero. With the expression given in (\ref{3.j}), it would be its {\it imaginary} part which provides us with the electric charge and the real part, the magnetic charge. Things would be reversed with the {\it dual} electromagnetic tensor
\beq
F_{ab}^\star= \frac12 e_{abcd} F^{cd} = -\ri \phi_{AB} \eps_{A' B'}+\ri \tilde{\eps}_{AB} \bar{\phi}_{A' B'}
\label{3.p}
\eeq
(see (\ref{3.k})), where it would now be the real part of the integral (\ref{3.g}) (with $ -\ri\phi_{AB}$ in place of $\phi_{AB}$) that provides us with the electric charge and the imaginary part, the magnetic charge (which would be zero by (\ref{3.n})).

 In an appropriately corresponding way, the source quantity for the spin-2 case of weak-field gravity, as arising from the 10 independent choices of the complex quantity $\ka^{AB}$, would not be expected to involve all the10 independent {\it complex} quantities that would arise from the 10-complex-dimensional space of solutions of (\ref{3.e}). Indeed, for a real space-time metric $g_{ab}$, there is a Hermiticity relation that reduces these 10 complex source quantities to 10 real ones, these comprising the 4 of mass-momentum and the 6 of angular momentum (see \cite{AT, PR}).

In the discussion which follows, we take our region $\mathcal V$ to be the spherically rotationally symmetric region $\mathcal R$ of section 2. Moreover, we shall be interested only in the choice of $\ka^{AB}$ that provides us with the {\it mass} of the source. This is singled out by the spherical rotational symmetry of the situation under consideration (together with spatial reflection symmetry, which rules out a ``magnetic mass'', i.e. NUT parameter \cite{NUT}, although this is in any case incompatible with normal globality considerations). Indeed, the entire region $\mathcal R$ under consideration here has been chosen to respect this spherical symmetry.

Although we are taking conformal flatness to be a good approximation to the metric of this space-time region $\mathcal R$ (being nowhere close to the large conformal curvature effects due to the mass concentration along the line $h$ with future end-point H), we need to bear in mind that the {\it metric} $g_{ab}$ of $\mathcal R$ is certainly {\it far} from being smooth in the neighbourhood of $\mathcal X$, so our considerations above must apply instead to small perturbations, within $\mathcal R$, of the (flat) metric $\tilde{g}_{ab}$ (see (\ref{2.pp})) of Minkowski space-time $\mathbb M$, given by
\beq
\rd \tilde{s}^2 = \rd\tau^2-\rd\Si^2 =\om^2 \rd s^2,\ {\rm where}\ \om=\pm F(\tau)^{-1}
\label{3.q}
\eeq
Accordingly, we use the notation $\tilde{\psi}_{ABCD}$ for the small perturbations away from the Minkowski metric $\tilde{g}_{ab}$ rather than the $\psi_{ABCD}$ of the above discussion. Moreover, as in Section 2, the covariant derivative operator $\tilde{\na}_a=\tilde{\na}_{AA' }$ is taken to be that of this (now flat) background. The region $\mathcal R$, with its $\tilde{g}_{ab}$ metric unites both regions $\mathcal A$ and $\mathcal B$ across their CCC-assumed conformally smooth crossover $\mathcal X$ (within $\mathcal R$) this being fundamental for our considerations here.

The nature of the region $\mathcal R$, whose conformal metric is taken to be a small perturbation away from the flat $\tilde{g}_{ab}$ metric of $\mathbb M$, allows us to regard $\mathcal R$ to be viewed as an actual spherically symmetric {\it subset} of $\mathbb M$, with a symmetry axis $\tilde{h}$, this being the $\tau$-axis of the metric expression (\ref{3.q}). This extension of $\mathcal R$ inwards to $\tilde{h}$ applies all down the cylinder $\mathcal C$, including negative values of $\tau$. However, his extension of $\mathcal R$ is to retain the physical metric $g_{ab}$ of $\mathcal M$, and its conformal structure will differ very greatly from $\mathbb M$’s flat $\tilde{g}_{ab}$-metric, as we move inwards towards to the world-line $h$ of the galactic cluster under consideration, and its resulting black hole. Nevertheless, we can identify the world-line $h$ with a portion of the axis $\tilde{h}$. The 3-plane $\tau=0$ in (\ref{3.q}) is taken to be the crossover 3-surface $\mathcal X$, when within the region $\mathcal R$, but it is not obvious that when we get close to the region of very large conformal curvature near the Hawking point H, that the origin of coordinates O for (\ref{3.q}) might not differ significantly, either to the past or future along $\tilde{h}$, from the actual Hawking point H.

What does this mean, when there is no canonical identification between the actual space-time $\mathcal M$ and the Minkowski space $\mathbb M$, when we are concerned with the region of $\mathcal M$’s large conformal curvature? The issue is of no real concern for our considerations here, except along the axis $h$ in $\mathcal M$ and $\tilde{h}$ in $\mathbb M$. Being the spherical symmetry axis in each case, the lines themselves can be identified with each other, but not necessarily pointwise. If we are concerned with ``actual points'' on the $h$ prior to H within the space-time $\mathcal M$ we have a problem, because for most of its latest history, just prior to H, there will be a black hole centred on this line, and the ``actual location'' of a point on this line is problematic. Accordingly, we shall be concerned only with the ``temporal location'' of the point H itself, for which we need consider only the situation within the aeon $\mathcal A$, its temporal location being identified by were its future light cone intersects $\mathcal C$, namely at $\mathcal T$ (see Section 2). This may be compared with the intersection with $\mathcal C$ of the future light cone of the origin point O in $\mathbb M$, since $\mathbb M$ is canonically identified with $\mathcal M$ within the region $\mathcal R$. Thus, we can raise the issue of whether or not the point O lies above or below H, namely whether or not O’s future light cone’s intersection with $\mathcal C$, in $\mathbb M$’s metric, is above or below $\mathcal T$.

The essential issue is whether or not the crossover 3-surface $\mathcal X$ remains, in an appropriate sense, ``flat'' when near a Hawking point H on $\mathcal X$. If, in this sense, $\mathcal X$ has a ``hill'' or a ``dip'' near H, we would find that O’s future light cone lies to the future or to the past of  $\mathcal T$. This could have an effect on the observed size of the Hawking points, their being a bit smaller in the case of a hill and a bit larger in the case of a dip. 

In view of the apparent anomaly, concerning this diameter, that will be addressed below, one might question whether a significant separation between H and O might be large enough to become relevant. However, this seems unlikely to us, because any such separation would have to come about from the actual mass value foreach galactic cluster, since there are no other parameters available in the very late stages of the aeon $\mathcal B$.  Moreover, there are vast variations in the sizes of galactic clusters, as observed in our aeon $\mathcal A$, and we must suppose that such large variations would have applied also to the earlier aeon also. However, the stark conclusion of \cite{AMNP}, at a 99.98\% confidence level, shows that the observed spots are all pretty-well of the same angular size, without the very large differences that would be anticipated if the separations between H and O for different Hawking points were to depend significantly on such mass differences between different Hawking points. Accordingly, we are rejecting an explanation of this kind for the apparent anomaly discussed in Section V, for this reason. It seems reasonable to suppose that there is, in fact, no significant separation between H and O, and in our discussion that follows, we take H and O to be effectively coincident.

We choose standard coordinates for $\mathbb M$ as in (3.q), where the time-axis ($\tau$-axis) is $\tilde{h}$, where its unit vector $t^a$ points along this axis. We take $x^a$ to represent the Minkowskian position vector of a variable point under consideration, with respect to O (=H), and we have $x^0=\tau$. We now define $\tilde{\ka}^{AB}$ by
\beq
\tilde{\ka}^{AB} = x^{(A}{}_{C'} t^{B)C'},
\label{3.r}
\eeq
for which the satisfaction of (\ref{3.e}) follows immediately from
\beq
\tilde{\na}_a x^b= \de_a^ b,\     {\rm i.e.} \tilde{\na}_{AA'} x^{BB'} = \tilde{\eps}_A{}^B \tilde{\eps}_{A'}{}^{B'}.
\label{3.s}
\eeq
This expression for $\tilde{\ka}^{AB}$ is appropriately spherically symmetrical (involving no directional quantities other than the time-axis vector $t^a$) so must necessarily provide us with a quantity $M$ proportional to the mass when $\tilde{\ka}^{AB}$ is substituted into the tilde versions of (\ref{3.f}) and (\ref{3.g}) providing:
\bea
M&=&\int \phi_{AB} \eps_{A' B'} \rd x^a \rd x^b  =\int \psi_{ABCD} \ka^{CD} \eps_{A' B'} \rd x^a \rd x^b \nn\\
&=&\int \psi_{ABCD} x^{(A}{}_{E'} t^{B) E' }  \eps_{A' B' } \rd x^a \rd x^b
\label{3.t}
\eea
At this stage, we can already be assured that $M$ must be conserved from aeon $\mathcal A$ to aeon $\mathcal B$, where we take $\mathcal S$ to be a $\tau=$const cross-section of the cylinder $\mathcal C$. This would follow if the equation (\ref{3.i}) holds, with $\tilde{\phi}_{AB}$ satisfying the free Maxwell equation (see (\ref{3.d}))
\beq
\tilde{\na}^{AA'} \tilde{\phi}_{AB}=0,
\label{3.u}
\eeq
which holds whenever $\tilde{\na}^{AA' } \tilde{\psi}_{ABCD}=0$. The latter equation follows from the vacuum equations (\ref{2.ff}) together with conformal invariance, in the late stages of aeon $\mathcal B$, but in the early stages of aeon $\mathcal A$ we need to appeal to a different consideration. This is that the integral (\ref{3.t}) is an expression defining the total ``mass-charge'' (magnetic or electric) of the $\tilde{\phi}_{AB}$ field surrounded by the surface $\mathcal S$ at conformal time $\tau$. This total ``mass-charge'' could change only if there is a local flux across a surface element $\rd \mathcal S$ of the 2-surface $\mathcal S$. However, whatever one’s expectations might be for the distribution of the matter in the very early universe, a local isotropy would normally be assumed, so that the ``mass-charge'' flux for $\tilde{\phi}_{AB}$ would be the same in both directions across $\mathcal S$. This is also anticipated on the basis of CCC, where conformal smoothness across $\mathcal X$ implies an initial local isotropy at the beginning of the ensuing aeon.

In order to be more specific about the quantity $M$ and to obtain the exact coefficient relating it to the actual value of the mass of a system, we can examine the case of the Schwarzschild solution for mass $m$, for which we have (see \cite{PR} pp 107, 110)
\beq
\tilde{\psi}_{ABCD}   =-\frac{6mG}{r^3}   o_{(A} o_{B} i_C i_{D)}
\label{3.xk}
\eeq
where $G$ is the gravitational constant, and where, at each point of $\mathbb M$  away from $\tilde{h}$ we introduce a spin frame $(o_A,i_A)$, with $o_A i^A=1$, where the flagpole (see \cite{PR}) of $o_A$ points in the outward null direction away from $\tilde{h}$ and the flagpole of $i_A$ in the inward null direction. As for the choice of flag planes, i.e. phases, of these spinors, this is unimportant since in all the expression of relevance these spinors appear in pairs where the phases cancel out. Also, for convenience of notation, we do not attach tildes to $o_A$  or $i_A$, despite the fact that we raise and lower their indices with $\tilde{g}_{ab}$ and associated epsilons (e.g. $o^A=\tilde{\eps}^{AB} o_B$, $i_{B'}=i^{A'} \tilde{\eps}_{A' B' }$,and  $\tilde{\eps}_{AB}=o_A i_B-i_A o_B)$.

In accordance with standard procedures relating spin frames with Minkowskian orthonormal frames (see \cite{PR, WP}), we can arrange the scalings of $o^A$ and  $i^A$ so that
\beq
t^a = \frac{1}{\sqrt{2}} (o^A o^{A'}+i^A i^{A' })
\label{3.xl}
\eeq
and at the point with position vector $x^a$
\beq
x^a =\frac{r}{\sqrt{2}} (o^A o^{A'}-i^A i^{A' }) + \tau t^a, 
\label{3.xm}
\eeq
where $r$ is the spatial $\mathbb M$-distance from $\tilde{h}$, and where $\tau$ is the conformal time parameter of (\ref{3.q}). Putting these two together, we obtain
\beq
\tilde{\phi}_{AB}=-\frac{6mG}{r^2}   o_{(A} o_B i_C i_{D)} o^C i^D
=\frac{mG}{r^2}   o_{(A} i_{B)}.
\label{3.xn}
\eeq
Taking $\mathcal S$ to be a sphere of radius $r$ in the Euclidean 3-space  $\tau=$const., we find that 
\beq
\int \phi_{AB} \eps_{A' B' } \rd x^a \rd x^b  = 4\pi mG
\label{3.xp}
\eeq
i.e.,
\beq
\int \psi_{ABCD} \ka^{CD} \eps_{A' B'} \rd x^a \rd x^b  = 4\pi mG
\eeq
Accordingly,
\beq
M= 4\pi mG.
\eeq

\vspace{5mm}
\section{Temperature profile within Hawking spots}

We can now use the theorem proven in the previous section to estimate the increase of temperature in the center of Hawking spots on the CMB map. We compare the injected energy (approximately equal to the mass of the largest galaxy clusters) to the average energy density at the last scattering surface. Assuming that the previous aeon was similar to ours similar scenario should have taken place then. On a  conformal diagram of our aeon an evaporating black hole in the previous aeon would look like  as a point-like injection of energy at the initial space-like scri of our aeon since the evaporation of large black holes of mass $M$ in Minkowski spacetime takes a very long time.
A Hawking spot was defined in \cite{AMNP} as a disk on the CMB map with an elevated temperature and having some special profile -- temperature decreasing from the center outwards. Using the theorem proven in the previous section we can estimate the additional contribution to the temperature at the time of the last scattering. We assume that the largest black holes evaporating in the previous aeon come from the largest gravitationally bound structures in the Universe i.e. the biggest galaxy clusters. Masses of the largest galaxy clusters in our Universe are estimated as
\beq
M_{gc}\sim \ 10^{15}\ M_\odot
\eeq  
so the whole process of evaporation, although extremely long in cosmic time, on the conformal diagram is effectively a point on the future space-like scri. We assume that energy coming from the evaporation of these largest structures (although the black holes of such gigantic masses are not yet present, see \cite{NT}, but are presumably created in a very remote future) is the source of the temperature elevation inside the Hawking spots on the CMB map in the next aeon. 

To calculate the temperature elevation on the CMB map on the Hawking spots we start from the density at the time of last scattering (see section V C):
\beq
\rho_{LS}\sim \frac{\rho_0}{a_{LS}^3} \sim 1.2\cdot 10^{-20}\ {\rm kg/m}^3
\eeq
The maximal average density of the injected mass (energy) from the evaporation of the largest galaxy clusters in the previous aeon is given by
\beq
\de\rho\sim \frac{3 M_{gc}}{4\pi(\eta^c_{LS})^3 c^3 a_{LS}^3}\sim 10^{-22}\ {\rm kg/m}^3
\eeq
where we used total conformal time $\eta^c$ from (\ref{etatotal}) and we used the theorem from sections II and III. Therefore  we get
\beq
\frac{\de T}{T}\sim\frac{\de \rho}{4\rho_{LS}}\sim 10^{-3}
\label{deltaT}
\eeq
where this theoretical estimate based on galactic clusters masses before the crossover is the same as the results of \cite{AMNP} where an observational estimate for the centers of the most significant Hawking spots on the Planck CMB map \cite{PMC} after the crossover is also $\frac{\de T}{T}\sim 10^{-3}$. Such a quantitative agreement points to the conclusion that the physics in the previous aeon was, at least in the gravitational sector, similar to ours.

\section{Cosmological Epochs}

In order to make a numerical estimates of the conformal times in different cosmological epochs we need to have a unified description of the scale factor as a function of the conformal time. 

The usual cosmological epochs are described by specifying the energy-momentum tensor, assuming the isotropic and homogeneous  metric (with vanishing Weyl tensor) and solving the Einstein equations (we use signature $(+,-,-,-)$ and put $c=1$)
\beq
R^{a}_{b}-\frac12R \de^{a}_{b}-\La \de^{a}_{b}=8\pi G T^a_b
\eeq
In standard cosmologu we distinguish three epochs
\bit
\item
Radiation Dominated Epoch (RDE):\\$T^a_b= \diag(\rho,-\rho/3,-\rho/3,-\rho/3)$, $\rho\gg \La$
\item
Matter Dominated Epoch (MDE):\\ $T^a_b= \diag(\rho,0,0,0)$, $\rho\gg \La$
\item
Lambda Dominated Epoch (LDE): $T^a_b=0$
\eit
and the Universe is described by the sequence
\beq
{\rm RDE}\to{\rm MDE}\to{\rm LDE}
\eeq
If we decompose the Riemann tensor as
\beq
\bR=\bC+\bE+\bS
\eeq
or in components ($C$ is the Weyl tensor)
\bea
\bR&=&R^{ab}{}_{cd}\nn\\
\bC&=&C^{ab}{}_{cd}\nn\\
\bE&=&E^{ab}{}_{cd}=2\tilde{R}^{[a}_{[c}\de^{b]}_{d]},\ \ \ \tilde{R}^a_c=R^a_c-\frac14 R\de^a_c\nn\\
\bS&=&\frac16 R\de^{[a}_{c}\de^{b]}_{d}
\eea
then $\bC=0$ in all three usual epochs and: $\bE\gg \bS$ in RDE,  $\bE\sim \bS$ in MDE and $\bE\ll \bS$ in LDE (in the Gravitational Wave Epoch discussed below $\bC\gg \bE,\bS$).

\vspace{5mm}
\subsection{Exact solution for the Radiation, Matter and Lambda Epochs}

In the usual cosmological evolution we assume that the metric takes the form
\beq
\rd s^2=a(\eta)^2\left(\rd\eta^2-\rd r^2-\frac{\sinh^2(2Kr)}{4K^2}\rd\Om^2\right)
\eeq
where $K$ can be real (spatially open Universe), pure imaginary (spatially closed Universe) or $K\to 0$ (spatially flat Universe), $\eta$ is the conformal time and $a(\eta)$ is the scale factor. We will give here a unified formula for the scale factor as a function of the conformal time for all three epochs.

The density is given by three components
\beq
8\pi G \rho(\eta)=\frac{3A^2}{a(\eta)^4}+\frac{12B^2}{a(\eta)^3}+3H_\La^2
\label{density}
\eeq
where $A$ corresponds to radiation, $B$ to matter and the cosmological constant $\La=3H_\La^2$. The cosmic time is given by the integral
\beq
t=\int a(\eta)\rd\eta
\eeq
Then the Friedmann equations read
\beq
\left(\frac{\rd a}{\rd\eta}\right)^2=A^2+4B^2a(\eta)+4K^2a(\eta)^2+H_\La^2a(\eta)^4
\label{FE}
\eeq
With the assumption that $A$, $B$ and $K$ are constant the general solution reads
\beq
a(\eta)=\frac{P-\frac{K^2}{4B^2}-\frac{A}{2B^2}P'}{(P-\frac{K^2}{4B^2})^2-\frac{A^2H_\La^2}{4B^4}}
\label{Unisolution}
\eeq
$P$ is the Weierstrass function
\beq
P(\eta):=\wp(B\eta,g_2,g_3)
\eeq
and 
\bea
g_2&=&\frac{A^2H_\La^2}{B^4}+\frac{3K^4}{8B^4}\nn\\
g_3&=&-\frac{H_\La^2}{B^2}-\frac{K^6}{8B^6}+\frac{A^2H_\La^2K^2}{2B^6}
\eea
It satisfies
\beq
\left(\frac{\rd P}{\rd\eta}\right)^2=B^2(4P^3-g_2 P-g_3)
\label{Weier}
\eeq
One should emphasize that the solution is not valid in the period when $A$ and $B$ are not constant i.e. before the nucleosynthesis and in the very far future of an aeon when the metric is determined by gravitational waves in the Gravitational Wave Epoch (discussed in the next section). It is also important that $\eta$ is {\it not} the conformal time that elapsed from the crossover, the latter we will denote as $\eta^c$, and therefore we cannot extend this solution down to $\eta=0$ (what would correspond to the conformal time of the crossover). The relation of $\eta$ to $\eta^c$ will be discussed in the next section.

\subsection{Series expansions of the solution}

The spatial curvature of our Universe is observationally very small so we now discuss the solution (\ref{Unisolution}) in the case $K=0$
\beq
a(\eta)=\frac{P-\frac{A}{2B^2}P'}{P^2-\frac{A^2H_\La^2}{4B^4}}
\label{arm}
\eeq
where $P$ is the Weierstrass function
\beq
P(\eta):=\wp\left(B\eta,\frac{A^2H_\La^2}{B^4},-\frac{H_\La^2}{B^2}\right)
\label{Uniflat}
\eeq
where $\eta$ is the shifted conformal time as discussed in (\ref{etatotal}).
The scale factor $a(\eta)$ in (\ref{arm}) vanishes for $\eta\to 0$ and goes to infinity for a finite conformal time $\eta_\infty$ determined by the equation
\beq
P(\eta_\infty)=\frac{AH_\La}{2B^2}
\label{etainfty}
\eeq
The equation
\beq
\left(\frac{\rd P}{\rd\eta}\right)^2=B^2\left(4P^3-\frac{A^2H_\La^2}{B^4}P+\frac{H_\La^2}{B^2}\right)
\label{Weierflat}
\eeq
gives
\beq
P'(\eta_\infty)=-H_\La, \ \ \ P''(\eta_\infty)=\frac{A^2H^2_\La}{B^2}
\eeq

For practical purposes it is convenient to provide the expansion of (\ref{Uniflat}) in powers of the conformal time. We divide the series expansions into the early time ones and the late time ones.The constants that approximately correspond to our Universe from the nucleosynthesis up to now are given by
\beq
A= 1.80\cdot 10^{-20}\ {\rm s}^{-1},\ \ B = 6.3\cdot 10^{-19}\ {\rm s}^{-1}, 
\label{cons}
\eeq
while the value of $H_\La$ 
\beq
H_\La=1.83\cdot 10^{-18}{\rm s}^{-1}
\label{HLa}
\eeq
corresponds to the value of the cosmological constant $\La\sim 1.00\cdot 10^{-35}$ s$^{-2}$.

An expansion of the solution for the scale factor in (\ref{arm}) that is sufficiently accurate for $\eta<10^{18}$ s reads
\beq
a(\eta)=A\eta+B^2\eta^2+\frac17 H_\La^2 A B^4\eta^7+\frac{1}{28} H_\La^2 B^6\eta^8+\ldots
\label{earlysola}
\eeq
and for the cosmic time
\beq
t(\eta)=\frac12 A\eta^2+\frac13 B^2\eta^3+\frac{1}{56} H_\La^2 A B^4\eta^8+\frac{1}{252} H_\La^2 B^6\eta^9+\ldots
\label{earlysolt}
\eeq

An expansion that is sufficiently accurate for $10^{18}\ {\rm s}<\eta<\eta_\infty$ is given by
\beq
a(\eta)=\frac{1}{H_\La(\eta_\infty-\eta)}-\frac12 B^2(\eta_\infty-\eta)^2
+\frac3{28} B^4 H_\La(\eta_\infty-\eta)^5-\frac1{56} B^6 H_\La^2(\eta_\infty-\eta)^8+\ldots\nn
\label{latesola}
\eeq
and for the cosmic time 
\beq
t(\eta)=-\frac{1}{H_\La}\ln\left(\frac{H_\La(\eta_\infty-\eta)}{2.114}\right)+\frac16 B^2(\eta_\infty-\eta)^3
-\frac1{56} B^4 H_\La(\eta_\infty-\eta)^6+\frac1{504} B^6 H_\La^2(\eta_\infty-\eta)^9+\ldots\nn
\label{latesolt}
\eeq
where
\beq
\eta_\infty\sim1.943\cdot 10^{18}\ {\rm s}
\label{etai}
\eeq

The important moments in the history of the Universe are:
\bit
\item
 the onset of nucleosynthesis ($T\sim 10^{10}$ K)
\beq
a_{NS}= 2.70\cdot 10^{-10},\ t_{NS} = 2.02\ {\rm s},\  \eta_{NS}=1.50\cdot 10^{10}\ {\rm s}
\label{NS}
\eeq
\item
radiation-matter transition ($a_{RM}=A^2/(4B^2)$)
\beq
a_{RM}= 2.04\cdot 10^{-4},\ \ t_{RM} = 9.05\cdot 10^{11}\ {\rm s},\ \ \eta_{RM}=9.4\cdot 10^{15}\ {\rm s}\nn
\label{RM}
\eeq
\item
Last Scattering (release of the CMB)
\beq
a_{LS}= 1/1100,\ \ t_{LS} = 1.17\cdot 10^{13}\ {\rm s},\ \eta_{LS}=3.0\cdot 10^{16}\ {\rm s}
\label{LS}
\eeq
\item
our present time ($a(\eta_0)=1$) 
\beq
\eta_0\sim 1.424\cdot 10^{18}\ {\rm s}, \ \ \rho_0\sim 9\cdot 10^{-27}\ {\rm kg/m}^3
\label{eta0}
\eeq
\eit
It is important to emphasize again that these values of $\eta$ do not correspond to the conformal time that elapsed from the crossover and that the solution cannot be extended to $\eta_\infty$ because of the Gravitational Wave Epoch at the end of the aeon (and possibly because of other unknown factors like eventual slow decay of Dark Matter).

\section{Gravitational Wave Epoch}

We propose that at the end of an aeon when approaching the crossover there is one more 'purely gravitational' epoch
Gravitational Wave Epoch (GWE) where there are large fluctuations of the Weyl tensor in the form of a 'gas' of gravitational waves and it extends through the crossover up to the beginning of the Radiation Epoch with its usual cosmological expansion afterwards. 

\subsection{GWE at the end of an aeon}
At the end of an aeon there are several possible sources of the gravitational waves in GWE like black holes evaporation, Dark Matter decay or the Cosmological Gravitational Background (CGB). There are several proposals of particles being the candidates for Dark Matter, for example axions or the recent proposal of Planck mass gravitinos \cite{MN1}, but they are stable (or almost stable) until they are 'swallowed' by the giant black holes that ultimately evaporate into gravitational waves. In \cite{PDM} there was a proposal that in CCC there is a possibility that Dark Matter consists of massive particles, 'erebons', interacting only gravitationally that slowly decay into gravitons. GWE  has large local fluctuations of the Weyl tensor (and small averaged Weyl tensor). In distinction to the usual 'metric' epochs (RDE, MDE and LDE) in the 'gravitational' epoch conformal weights are different than in the 'metric' epochs (these weights are not 'canonical'). The Einstein equations do not carry full information about this 'gravitational' epoch and we have to use the spinorial description as in (\ref{2.f}). We propose that at the end of the aeon, after the evaporation of black holes, we enter the GWE followed by the crossover. The random local fluctuations of the Weyl tensor in GWE have a net positive focusing effect on light rays coming about from the non-linear {\it astigmatic} focusing arising from Weyl curvature (see \cite{PR}). Another proposal of the source of GWE at the end of an aeon is that the large fluctuations of the Weyl tensor could emerge  just before the crossover from the Cosmic  Gravitational Background (CGB) coming from many uncorrelated sources and permeating the whole Universe (review of the CGB and the possibility of its detection is given for example in \cite{RC}). The CGB has presumably the parameters
\beq
h\sim 10^{-16},\ \ \  k\sim 10^{-8}\ {\rm Hz}
\label{CGBvalues}
\eeq
where $h$ is the amplitude and $k$ is the frequency. It is unfortunately impossible to measure CGB at present because of its very low frequency. However, the presence of the cosmological constant and the exponential expansion can increase the importance of this background and mark the beginning of the GW Epoch.

To estimate heuristically when such a beginning of GWE can take place we consider one spherical gravitational wave (of strain $h$, $h\ll 1$) in the presence of the cosmological constant $\La=3 H_\La^2$:
\beq\label{gravwaveh}
\rd s^2=\frac{1}{(H_\La(\eti-\eta))^2}\left(\rd\eta^2-\rd r^2-r^2\rd\Om+2hrF(\eta)\sin(\th)\cos(kr)\rd\th\rd\phi\right)
\eeq
As discussed in the previous sections $\eti$ is the conformal time corresponding to the crossover.
Imposing the condition that the presence of the gravitational wave does not contribute to the Ricci tensor (but only to the Weyl tensor) we get the equation at the first order in $h$:
\beq
F''(\eta)+F(\eta)\left(k^2-\frac{8}{(\eti-\eta)^2}\right)=0
\eeq
with the solution in terms of Bessel functions
\beq
F(\eta)=(\eti-\eta)^{\frac12}\left( c_1 J_{\frac{\sqrt{33}}{2}}(k(\eti-\eta))+c_2Y_{\frac{\sqrt{33}}{2}}(k(\eti-\eta))\right)
\eeq
For $k(\eti-\eta)\gg 1$ the wave is of the expected form $\cos(k(\eta-r))/r$  and $\sin(k(\eta-r))/r$. However, when $\eta$ approaches the end of the aeon, $k(\eti-\eta)\to 0$, the $Y$ function goes to infinity and the gravitational waves start to have an important contribution to the metric tensor and the Weyl tensor. We can estimate, using (\ref{CGBvalues}) and the properties of the Bessel $Y$ function:
\beq
h(k(\eti-\eta_{GW}))^{-\sqrt{33}/2+1/2}>  1\Ra (\eti-\eta_{GW})<10\ {\rm s}
\eeq
At such late times the ansatz (\ref{gravwaveh}) is no longer appropriate and the fluctuations of the Weyl tensor are large.
Using (\ref{HLa})
\beq
H_\La=1.8451\cdot 10^{-18}{\rm s}^{-1}
\eeq
we get the scale factor when the gravitational waves start to be important
\beq
a_{GW}\sim(H_\La (\eti-\eta_{GW}))^{-1}\sim 6\cdot 10^{16}
\eeq
Therefore the onset of the Gravitational Wave Epoch starts at the cosmic time
\beq
t_{GW}-t_0\sim H_\La^{-1}\ln a_{GW}\sim 2\cdot 10^{19}\ {\rm s}\sim 10^{3}\ {\rm bln\ y}
\eeq
The 'gas' of gravitational waves should satisfy the same condition at order $h$ -- when $h$ is small the sum of solutions is still a solution. However at higher orders in $h$, because of nonlinearity of gravity, one would need to have a different description of the 'gas' of gravitational waves in the Gravitational Wave Epoch. A proposal of the equations governing the period both before and after the crossover in CCC is given in \cite{P2,PT}.

As a side remark one can give a example describing the approximate LIGO/Virgo/Kagra \cite{LVK} gravitational plane-wave. A good approximation of the amplitude (travelling in the $z$ direction with the speed of light) at the detector placed at $z$ and just before the observed black holes' merger is given by
\beq
A\sim h\frac{\sin\left((k(z/c-t)^{\frac23}\right)}{\sqrt{k(z/c-t)}},\ \ \ \ t<\frac{z}{c}
\eeq
where the standard values observed by LIGO/Virgo/Kagra \cite{LVK} $h\sim 10^{-21}$ and $k\sim 100$ Hz ($t>z/c$ would correspond to the ringdown). An ansatz for the metric (for one of the polarizations travelling in the $z$ direction)) is
\beq\label{gravwavesolution}
\rd s^2=c^2\rd t^2-\rd x^2-\rd y^2-\rd z^2
+2h\frac{\sin(k(z/c-t)^\frac23)}{\sqrt{k(z/c-t)}}\rd x \rd y
-h^2 p(k(z/c-t))\rd x^2
\eeq
Requiring that
\beq
R_{\mu\nu}=O(h^3)
\eeq
gives a differential equation for the function $p(s)$ that can be solved (with $p(0)=0$):
\beq
p(s)=s^\frac13+\frac{7\left(1-\cos(2 s^\frac23)\right)}{16s}+\frac{5\sin(2 s^\frac23)}{16s^\frac13}
-\frac{2\sqrt{\pi}}{3}{\rm C}\left(\frac{2s^\frac13}{\sqrt{\pi}}\right)-\frac{s}{12}\left({\rm si}(2s^\frac23)
+\frac{\cos(2s^\frac23)}{2s^\frac23}\right)
\eeq
where C$(s)=\int_0^s\rd t\cos(t^2)$is the Fresnel C function and si$(s)=-\int_s^\infty\rd t\frac{\sin(t)}{t}$ is the integral sine function.
Weyl tensor for the metric (\ref{gravwavesolution}) is of order $h$
\beq
C_{txty}=\frac29\frac{(-27/16 + s^{1/3})s\sin(s^{2/3}) + 2s^{2/3}\cos(s^{2/3})}{s^{5/2}}hk^2 +O(h^2)
\eeq
where $s=k(z/c-t)$. Therefore the solution (\ref{gravwavesolution}) describes a genuine gravitational planewave with vanishing Ricci tensor up to $O(h^3)$ ($R_{\mu\nu}=O(h^4)$ or higher and would require further corrections to (\ref{gravwavesolution}).

\subsection{GWE after the crossover and the total conformal time}

We start the discussion of the GWE just after the crossover, i.e. at the beginning of a new aeon, from the fact that the conformal time $\eta$ used in (\ref{Unisolution}) is not the total conformal time elapsed from the crossover \cite{AG}. The total conformal time elapsed from the crossover we denote by $\eta^c$ where 
\beq
\eta^c=\eta+\eta^G
\eeq
and $\eta^G$, is the conformal time elapsed from the crossover to the moment $\eta=0$ as defined in the equation (\ref{arm}).The conformal time $\eta^G$ is to be determined from observations. In \cite{AMNP} it was established that the angular radius of the Hawking spots is around (see also \cite{CCG} table 3)
\beq
r_{HS}\sim 0.03\div 0.04
\label{robs}
\eeq
Requiring that 
\beq
\frac{\eta^c_{LS}}{\eta^c_0}=\frac{\eta_{LS}+\eta^G}{\eta_0+\eta^G}=r_{HS}\sim 0.03\div 0.04
\eeq
where $\eta_{LS}$ is given by (\ref{LS}) and $\eta_{0}$ is given by (\ref{eta0}), we get
\beq
\eta^G\sim 2.0\cdot 10^{16}\ {\rm s}
\label{etaren}
\eeq
Therefore the total conformal times elapsed from the crossover are equal to
\bea
\eta^c_{NS}&=&\eta_{NS}+\eta^G\sim 2.0 \cdot 10^{16}\  {\rm s},\nn\\
\eta^c_{LS}&=&\eta_{LS}+\eta^G\sim 4.9 \cdot 10^{16}\  {\rm s},\nn\\
\eta^c_0&=&\eta_0+\eta^G\sim 1.48 \cdot 10^{18}\ {\rm s}
\label{etatotal}
\eea
with the scale factors given by (\ref{NS}), (\ref{LS}) and (\ref{eta0}), respectively.

A possible explanation of the value of $\eta^G$ would be to assume that just after the crossover we have the GWE that continues from the end of the previous aeon. The dominance of the gravitational waves in this very early time gives the epoch very different from the usual cosmological epochs and the slow transition into the Radiation Epoch is difficult to describe in detail -- the equations are proposed in \cite{P2,PT} but they are very difficult to solve with large fluctuations of the  Weyl tensor. One can expect that these equations should lead on one hand in the effective description to the proper $\eta^G$ at the beginning of the Radiation Era and on the other to go smoothly smoothly to the known 
\beq
a(\eta)=A\eta+B^2\eta^2
\eeq
for times later than the nucleosynthesis where $A$ and $B$ are given by (\ref{cons}). The presence of $\eta^G$ at the end of GWE can provide the explanation of the observational fact that the causally connected region starting at the crossover and ending at the last scattering surface is bigger than naively expected.  

\section{Negative spatial curvature in CCC}

It is important to note that CCC predicts (requires?) negative spatial curvature of the Universe. The main ingredient of the argument is the consideration of the future light cone of a point $p$, which we can think of being on a crossover surface, though this is not essential. As this cone expands it will encounter various regions of mater, whose energy-momentum tensor provides a bit of positive focussing, and perhaps other places where it encounters gravitational fields which causes astigmatic focussing which, due to no-linear effects will ultimately provide a bit more positive focussing. There is no negative focussing, because, classically, there is no effective negative focussing. All this can be pretty small, and certainly not sufficient, generally, to turn round the outward expansion of $p$'s future light cone. If we make the assumption that the overall spatial geometry of the universe is non-negative, as we follow this cone through aeon after aeon, the local divergence of the rays becomes smaller and smaller until this divergence becomes tiny enough that even the relatively tiny inward focussing becomes enough to convert the outward local divergence of the rays to become inward, so that the light cone starts to obtain smaller and smaller local spatial cross-sections, and once the rays begin to converge there is nothing to stop them crumpling inwards until their spatial cross-sections reduce to zero. This conclusion would be bad news for CCC, if only because it contradicts the de-Sitter-like future of an aeon, for which the expansion should continue indefinitely. Therefore, the inevitable conclusion is the falsity of the non-negative nature of the spatial curvature. With negative spatial curvature, the expansion of the rays does not slow down to zero (or less), but it continues to increase into the future, so there is no problem of the kind explained above. It would be interesting to see how the observational limits on negativity of the spatial curvature might relate to the amount of positive focussing that we expect, for the explicit positive focussing effects that we consider in our paper. These limits were provided by the Planck data \cite{PMC} where 
\beq
- (H^2 r^2 a^2)^{-1}\leqslant 0.0007 (\pm 0.0019)
\eeq
(assuming that $r^2$ is negative) so the negative spatial curvature is indeed extremely tiny.

\section{Magnetic fields and B-modes}

The presence of the so called B-modes on the CMB map is usually associated with the division of the Weyl tensor into the E-modes and B-modes. However, the presence of the usual magnetic field would make the appearance of these B-modes much more natural. We now calculate how large is the magnetic field contribution to the energy of the Hawking spots.

We start from the estimate of the energy stored in the magnetic field within the largest galactic clusters. We assume that the average value of the field is $\sim 1\ \mu$G i.e. $10^{-10}$ T and the size $\sim$ 10 mln ly $\sim$ $10^{23}$ m. Therefore the energy stored in the magnetic field within the galactic cluster
\beq
E_B=\frac{1}{2\mu_0}B^2\,V\sim 10^{55}\ {\rm J}
\eeq
On the other hand the energy stored in the mass of the largest galactic clusters ($M\sim 10^{15}\, M_\odot$) can be estimated as
\beq
E_M\sim Mc^2\sim 10^{62}\ {\rm J}
\eeq
Therefore, the temperature fluctuation coming from the magnetic field within the Hawking spot can be estimated as 
\beq
\frac{\de T_B}{T}\sim \frac{E_B}{E_M}\cdot \frac{\de T}{T}\sim10^{-7}\cdot 10^{-3}=10^{-10}
\eeq
i.e. around a nK (we used (\ref{deltaT}) for the fluctuation of temperature within the Hawking spot). The presence of such fluctuations could have a measurable effect creating the B-modes on the CMB map.

\vspace{5mm}
\section{Grand Arc and Big Ring}

In \cite{AML} and \cite{AML2}  it was shown that there are very big structures on the sky (Grand Arc and Big Ring). Such structures are very massive and do not change much on the conformal diagram so we can project them vertically to the initial singularity and compare the size with the time they could have been born in the previous aeon. 

The Grand Arc extends $\eta_{GA}\sim 10^{17}$ s (1 Gpc, present size) at $z\sim 0.8$. If we assume that the structure was born by a collision of massive black holes in the previous aeon and that the physics in the previous aeon was similar to ours then the collision should have taken place at ($\eta_\infty= 2.00\cdot 10^{18}$ s)
\beq
\eta_{col}\sim \eta_\infty -\eta_{GA}\sim 1.9\cdot 10^{18}\ {\rm s}
\eeq
what corresponds to (see (\ref{latesolt}))
\beq
t_{col}\sim 42\cdot 10^9\ {\rm years}
\eeq
i.e.at  95\% of the conformal diagram and 3 times the age of our present Universe in the previous aeon. If we assume that the Grand Arc is part of the full ring of radius , say,  $\eta_{GA}\sim 3\cdot 10^{17}$ s (3 Gpc) then the collision time would be earlier
\beq
\eta_{col}\sim \eta_\infty -\eta_{GA}\sim 1.7\cdot 10^{18}\ {\rm s}\Ra 
t_{col}\sim 23\cdot 10^9\ {\rm years}
\eeq
i.e. at  85\% of the conformal diagram and twice the age our present Universe (but in the previous aeon). Both numbers seem to be plausible for the collisions of very large black holes since in our Universe even now there exist huge black holes of the order of $10^{11}$ solar masses or more.

 The Big Ring is smaller \cite{AML2} and has radius $\eta_{BR}\sim 2\cdot 10^{16}$ s (200 Mpc, present size) at the same redshift $z\sim 0.8$. If we assume that the structure was born by a collision of massive black holes in the previous aeon then the collision should have taken place at
\beq
\eta_{col}\sim \eta_\infty -\eta_{BR}\sim 1.98\cdot 10^{18}\ {\rm s}
\eeq
what corresponds to (see (\ref{latesolt}))
\beq
t_{col}\sim 70\cdot 10^9\ {\rm years}
\eeq
i.e. at  99\% of the conformal diagram and 5 times the age of our present Universe (in the previous aeon) what again seems to be plausible for the collisions of very large black holes at the end of an aeon.

\section{Conclusions}

 In the paper a theorem is proved, using twistor theory introduced by one of the authors, that within the Cyclic Conformal Cosmology there actually exists an invariant that allows to relate the energy density in the previous aeon to the energy density in the subsequent aeon after the conformal transformation. Using this invariant it is shown that, surprisingly, there is a {\it quantitative} agreement between the masses of the black holes from the largest galaxy clusters (approximately $10^{15}$ solar masses) in the previous aeon and the raise of the temperature in the Hawking spots on the CMB map of the next aeon coming from the evaporation of these black holes. This agreement suggests that physics in the previous aeon was either identical or similar to the one we have in our aeon, at least in the gravitational sector. The scenario is proposed where the gravitational waves constitute the main ingredient of very late Universe, a Gravitational Wave Epoch with large local fluctuations of the Weyl tensor in presence of the cosmological constant, followed by a crossover and the continuation of the GWE at the beginning of the next aeon.  
 
\vspace{0.8cm}
\noindent
 {\bf Acknowledgments:} 
 R.P. is grateful to John Moussouris for financial assistance. K.A.M. was partially supported by the Polish National Science Center grant UMO-2020/39/B/ST2/01279.


\begin{thebibliography}{99}
\bibitem{Killing} We see the considerable advantage, here, in the use of the 2-spinor formalism. In conventional tensor notation one would need to encode $\ka_{AB}$ into a the much more complicated-looking skew tensor $S_{ab} =\ka_{AB} \eps_{A' B' }+\eps_{AB} \bar{\ka}_{A' B'}$, subject to
$$
\na_{(a} S_{b)c}-\na_{(a} S_{c)b}+g_{a[b} \na^d S_{c]d}=0.
$$
Moreover, the key role of $S_{ab}$ as a spin-lowering operator is far from transparent in this tensor form. See \cite{PR}, eqn.6.4.6
\end{thebibliography}
\end{document}